\documentclass[twocolappendix,numberedappendix]{emulateapj}
\usepackage{amsmath}
\usepackage{graphicx,epstopdf}
\usepackage{multirow}
\usepackage{color}
\usepackage{latexsym}
\usepackage{amssymb}
\usepackage{epsfig}
\usepackage{verbatim}

\usepackage[colorlinks=true,linkcolor=blue,citecolor=blue]{hyperref}

\begin{document}

\defcitealias{GS2015}{Paper~I}

\title{Extended Heat Deposition in Hot Jupiters: Application to Ohmic Heating}

\author{Sivan Ginzburg and Re'em Sari}

\affil{Racah Institute of Physics, The Hebrew University, Jerusalem 91904, Israel}

\begin{abstract}
Many giant exoplanets in close orbits have observed radii which exceed theoretical predictions. One suggested explanation for this discrepancy is heat deposited deep inside the atmospheres of these ``hot Jupiters''. Here, we study extended power sources which distribute heat from the photosphere to the deep interior of the planet. Our analytical treatment is a generalization of a previous analysis of localized ``point sources''. We model the deposition profile as a power law in the optical depth and find that planetary cooling and contraction halt when the internal luminosity (i.e. cooling rate) of the planet drops below the heat deposited in the planet's convective region.
A slowdown in the evolutionary cooling prior to equilibrium is possible only for sources which do not extend to the planet's center. 
We estimate the Ohmic dissipation resulting from the interaction between the atmospheric winds and the planet's magnetic field, and apply our analytical model to Ohmically heated planets. Our model can account for the observed radii of most inflated planets which have equilibrium temperatures $\approx 1500\textrm{ K}-2500\textrm{ K}$, and are inflated to a radius $\approx 1.6 R_J$. However, some extremely inflated planets remain unexplained by our model. 
We also argue that Ohmically inflated planets have already reached their equilibrium phase, and no longer contract. Following Wu \& Lithwick who argued that Ohmic heating could only suspend and not reverse contraction, we calculate the time it takes Ohmic heating to re-inflate a cold planet to its equilibrium configuration.
We find that while it is possible to re-inflate a cold planet, the re-inflation timescales are longer by a factor of $\approx 30$ than the cooling time.
\end{abstract}


\section{Introduction}
\label{sec:introduction}

The discovery of extra-solar planets in the last two decades has been accompanied with a variety of surprises, which challenge standard planetary formation and evolution theories that were originally inspired by our solar system. One of these mysteries is the detection of close-orbit planets with radii as large as $\sim 2 R_J$, where $R_J$ is the radius of Jupiter \citep{Baraffe2010,Anderson2011,Chan2011,Hartman2011,SpiegelBurrows2013}. 
Theoretical evolution models predict that isolated gas giants older than $\sim 1 \textrm{ Gyr}$ cool and contract to a radius of about $1.0 R_J$ \citep{Burrows97,GS2015}. The proximity of the observed planets to their parent stars imposes a strong stellar irradiation that induces a deep radiative envelope at the outer edge of the otherwise fully convective planets \citep{Guillot96,ArrasBildsten2006}. This radiative layer slows down the evolutionary cooling of the planet compared with an isolated one, resulting in a higher bulk entropy and radius at a given age \citep{Burrows2000,Chabrier2004,ArrasBildsten2006,SpiegelBurrows2012,MarleauCumming2014}.  
However, at least some of the observed hot Jupiters have radii which exceed the theoretical predictions, even with stellar irradiation taken into account \citep{Baraffe2003,Burrows2007,Liu2008}. 

There have been several suggested explanations to the radius discrepancy \citep[see][for comprehensive reviews]{Baraffe2010,Baraffe2014,FortneyNettelmann2010,SpiegelBurrows2013}, varying from enhanced atmospheric opacities \citep{Burrows2007}, suppression of convective
heat loss \citep[][]{ChabrierBaraffe2007,LeconteChabrier2012}, or an extra power source inside the planet's atmosphere. In this work we focus on the effects of an extra power source in the atmosphere. Possible heat sources include tidal dissipation due to orbital eccentricity \citep{Bodenheimer2001,Bodenheimer2003,Gu2003,WinnHolman2005,Jackson2008,Liu2008,IbguiBurrows2009,Miller2009,Ibgui2010,Ibgui2011,Leconte2010},
``thermal tides'' \citep{ArrasSocrates2009a,ArrasSocrates2009b,ArrasSocrates2010,Socrates2013},
Ohmic heating \citep{BatyginStevenson2010,Perna2010,Perna2012,Batygin2011,HuangCumming2012,RauscherMenou2013,WuLithwick2013,RogersShowman2014}, turbulent mixing \citep{YoudinMitchell2010}, and dissipation of kinetic energy of the atmospheric circulation \citep{GuillotShowman2002,ShowmanGuillot2002}.
 
The influence of the additional heat on the planet's cooling history (and therefore on its radius) increases with its deposition depth inside the atmosphere \citep{GuillotShowman2002,Baraffe2003,WuLithwick2013}. In a previous work \citepalias[Ginzburg \& Sari 2015; hereafter][]{GS2015} we gave an intuitive analytic description of the effects of additional power sources on hot Jupiters, which explains this result and reproduces the numerical survey of \citet{SpiegelBurrows2013}. However, \citetalias{GS2015} and previous numerical works focus on the specific case of localized ``point-source'' energy deposition. In the current work we study a more general scenario, in which the deposited heat is distributed over a range of depths in the atmosphere \citep[see, e.g.,][]{Batygin2011}.

The outline of the paper is as follows. We summarize the results of \citetalias{GS2015} for localized power sources in Section \ref{sec:point}, and generalize them to extended sources in Section \ref{sec:power_law}. In Section \ref{sec:physical} we apply our model to the specific case of Ohmic dissipation, and quantitatively estimate the radii of Ohmically heated hot Jupiters, with a comparison to observations. In Section \ref{sec:reinflation} we discuss the possibility of re-inflating a cold planet. Our main conclusions are summarized in Section \ref{sec:conclusions}.

\section{Point Source Energy Deposition}
\label{sec:point}

In \citetalias{GS2015} we analyzed the effects of localized heat sources deep in the atmosphere on the cooling rate of irradiated gas giants. We summarize the analysis here because a similar technique is used in Section \ref{sec:power_law}. Our model is one-dimensional and does not differentiate between the day
and night sides of the planet \citep[see][]{SpiegelBurrows2013}. We first discuss the effects of stellar irradiation, and then incorporate an additional heat source.

\subsection{Irradiated Planets}
\label{subsec:irradiation}

Irradiated planets, in contrast with isolated ones, develop a deep radiative outer layer which governs the convective heat loss rate (i.e., internal luminosity) of the interior \citep{Guillot96,ArrasBildsten2006}. The radiative-convective transition and the internal luminosity are best analyzed in the $[\tau, U]$ plane, with
\begin{equation}\label{eq:tau}
\tau(r)=\int_r^\infty{\kappa\rho dr'}
\end{equation}
denoting the optical depth at radius $r$, and $U\equiv a_{\rm rad}T^4$ is the radiative energy density. The density, temperature, and opacity are denoted by $\rho$, $T$, and $\kappa$, respectively, and $a_{\rm rad}$ is the radiation constant.

Assuming power-law opacities $\kappa\propto\rho^a T^b$ \citep[see, e.g.,][]{ArrasBildsten2006,YoudinMitchell2010}, the convective interior is described by
\begin{equation}\label{eq:U_con}
\frac{U}{U_c}=\left(\frac{\tau}{\tau_c}\right)^\beta,
\end{equation} 
with $U_c\equiv a_{\rm rad} T_c^4 $ denoting the central radiation energy density, determined by the central temperature $T_c$, $\tau_c\sim\kappa_c\rho_c R$, with $\kappa_c$ denoting the estimate for the central opacity, if the power-law opacity could be extrapolated to the center, $\rho_c$ denoting the central density, and $R$ is the planet's radius. The power $\beta$ is given by
\begin{equation}\label{eq:beta}
\beta=\frac{4}{b+1+n(a+1)},
\end{equation}
where $n$ is the polytropic index.

The radiative envelope is characterized by the equilibrium temperature on the planet surface (i.e., photosphere)
\begin{equation}\label{eq:Teq}
T_{\rm eq}=(1-A)^{1/4} T_\sun\left(\frac{R_\sun}{2D}\right)^{1/2},
\end{equation}
with $T_\sun$ and $R_\sun$ denoting the stellar temperature and radius, respectively, and where $D$ and $A$ are the planet's orbital distance and albedo, respectively \citep[see, e.g.,][]{Guillot96}. This equilibrium temperature defines an energy density of $U_{\rm eq}\equiv a_{\rm rad}T_{\rm eq}^4$, and luminosity $L_{\rm eq}\equiv 4\pi R^2\sigma_{\rm SB} T_{\rm eq}^4$, where $\sigma_{\rm SB}$ is the Stefan-Boltzmann constant. The radiative profile is given by the diffusion approximation (valid for $\tau\gg 1$)
\begin{equation}\label{eq:U_rad}
U=U_{\rm eq}+\frac{3}{c}\frac{L_{\rm int}}{4\pi R^2}\tau,
\end{equation}
where $c$ is the speed of light and $L_{\rm int}$ is the internal luminosity.

According to the Schwarzschild criterion, convective instability develops when the radiative temperature profile is steeper than the adiabatic one. By differentiating Equations \eqref{eq:U_con} and \eqref{eq:U_rad} we find that a profile with $\beta>1$ is fully radiative, while for $\beta<1$ \citepalias[which is the relevant scenario; see][]{GS2015} convection sets in at the radiative-convective boundary, located at an optical depth of
\begin{equation}\label{eq:irradiated}
\tau_{\rm rad}\sim\frac{L_{\rm eq}}{L_{\rm int}}\sim\tau_c\left(\frac{U_{\rm eq}}{U_c}\right)^{1/\beta},
\end{equation}
where the radiative energy density is $U_{\rm eq}/(1-\beta)$. Equation \eqref{eq:irradiated} shows that increasing stellar irradiation decreases the internal luminosity and deepens the penetration of the radiative layer, which is isothermal to within a factor of $(1-\beta)^{-1/4}$ \citep[see also][]{Guillot96,Burrows2000,ArrasBildsten2006,YoudinMitchell2010,SpiegelBurrows2013}.
These results can be also obtained graphically by drawing a radiative tangent with a slope $3L_{\rm int}/(4\pi R^2 c)$ from the point $[0,U_{\rm eq}]$ to the convective profile \citepalias[see][]{GS2015}.
       
\subsection{Power Deposition}
\label{subsec:deposition}

In \citetalias{GS2015} we parametrized an energy point-source with its power $L_{\rm dep}$ and some deposition optical depth $\tau_{\rm dep}$. This source alters the radiative profile of Equation \eqref{eq:U_rad}
\begin{equation}\label{eq:deposit_rad}
\frac{dU}{d\tau}=\frac{3}{4\pi R^2c}\cdot
\begin{cases}
L_{\rm tot}\equiv L_{\rm int}+L_{\rm dep} & \tau<\tau_{\rm dep}
\\[1.5ex]
L_{\rm int} & \tau>\tau_{\rm dep}
\end{cases},
\end{equation}
with $L_{\rm tot}$ denoting the total luminosity for $\tau<\tau_{\rm dep}$. 
If we focus on intense deposition $L_{\rm dep}\gg L_{\rm int}$, then we may approximate $L_{\rm tot}\approx L_{\rm dep}$, and by analogy with Equation \eqref{eq:irradiated}, convection sets in at $\tau_b\sim L_{\rm eq}/L_{\rm dep}$ (we reserve the notation $\tau_{\rm rad}$ to the inner radiative-convective transition, as discussed below). In the regime $L_{\rm dep}\tau_{\rm dep}/L_{\rm eq}\gtrsim 1$ \citepalias[necessary for a significant effect, see][]{GS2015} a convective layer appears between $\tau_b$ and $\tau_{\rm dep}$, which we distinguish as the secondary convective region. The internal luminosity is found by drawing a radiative tangent from $[\tau_{\rm dep},U(\tau_{\rm dep})]$ to the main interior convective profile, with the transition point denoted by $\tau_{\rm rad}$. This is equivalent to drawing a tangent from $[0,U_{\rm iso}]$, with
\begin{equation}\label{eq:U_eq_eff_big}
\frac{U_{\rm iso}}{U_{\rm eq}}\approx\frac{1}{1-\beta}\frac{U(\tau_{\rm dep})}{U(\tau_b)}\sim\left(\frac{L_{\rm dep}\tau_{\rm dep}}{L_{\rm eq}}\right)^\beta,
\end{equation}
where $U(\tau_{\rm dep})$ is adiabatically related to $U(\tau_b)$. Therefore, the results of Section \ref{subsec:irradiation} are reproduced, but with $U_{\rm iso}$ instead of $U_{\rm eq}$ (note the change in notation of the deep isotherm from $U_{\rm eq}^{\rm eff}$ in \citetalias{GS2015} to $U_{\rm iso}$ here). Combining Equations \eqref{eq:irradiated} and \eqref{eq:U_eq_eff_big} shows that the internal luminosity is reduced from a value of $L_{\rm int}^0$ without heat sources to
\begin{equation}\label{eq:short}
\frac{L_{\rm int}}{L_{\rm int}^0}\sim\left(1+\frac{L_{\rm dep}\tau_{\rm dep}}{L_{\rm eq}}\right)^{-(1-\beta)},
\end{equation}
where we have interpolated with the weak heating regime ($L_{\rm dep}\tau_{\rm dep}/L_{\rm eq}\lesssim 1$). The conclusion is that additional heat sources slow the cooling rate if deposited deep enough.

\subsection{Effect of Heating on Planet Radius}
\label{subsec:radius}

The evolution of a planet's central temperature with time is determined by its internal luminosity: 
\begin{equation}\label{eq:cooling}
L_{\rm int}\sim-k_{\rm B}\frac{M}{m_p}\frac{dT_c}{dt},
\end{equation}
where $k_{\rm B}$ is the Boltzmann constant, $m_p$ is the proton mass, $M$ is the mass of the planet, and $t$ denotes time.

The radius of a planet can be determined directly by its central temperature.
The relation between the radius increase relative to the zero-temperature radius $\Delta R\approx R-0.9R_J$ and the central temperature can be derived either by using a linear approximation for $\Delta R\ll R_J$ 
\begin{equation}\label{eq:delta_r_t}
\Delta R\sim\frac{k_{\rm B}T_c}{m_pg},
\end{equation}
with $g\approx 10^3\textrm{ cm}\textrm{ s}^{-2}$ denoting the surface gravity \citep{ArrasBildsten2006,GS2015}, or by using a numerical radius-central-temperature curve \citep[e.g.,][]{Burrows97}.

At an age of $\sim 1\textrm{ Gyr}$, isolated Jupiter-mass planets reach a radius of $R\approx 1.0 R_J$ \citep{Burrows97}.
As explained in Section \ref{subsec:irradiation}, stellar irradiation slows down the planetary contraction, allowing more inflated planets at the same age. Specifically, hot Jupiters irradiated by an equilibrium temperature of $T_{\rm eq}\approx 1500\textrm{ K}$ are expected to reach a radius of $R\approx 1.3 R_J$ at this age \citep[see, e.g.,][]{Burrows2007,Liu2008,GS2015}.

As explained in Section \ref{subsec:deposition}, deep heat deposition slows down the cooling rate even more, resulting in an additional radius inflation at a given age. Moreover, deep deposition raises the final (equilibrium) central temperature of the planet (which is roughly equal to the temperature at the inner radiative-convective boundary; see Appendix \ref{sec:other_cases}, specifically Figure \ref{fig:scheme_stages}) from $T_{\rm eq}$ to $T_{\rm iso}\equiv (U_{\rm iso}/a_{\rm rad})^{1/4}$, given by Equation \eqref{eq:U_eq_eff_big}. When the planet reaches this equilibrium temperature, cooling and contraction stop entirely, and an enlarged equilibrium radius is retained. 

\section{Power-Law Energy Deposition}
\label{sec:power_law}

We now generalize the results of Section \ref{sec:point} to account for an extended source that spans a broad range in optical depth
\begin{equation}\label{eq:power_dep}
L_{\rm dep}(\tau)=\epsilon L_{\rm eq}\tau^{-\alpha}\qquad 1\leq\tau\leq\min{(\tau_{\rm cut},\tau_c}),
\end{equation} 
where $L_{\rm dep}(\tau)$ denotes the heat deposited deeper than $\tau$, and $\epsilon$ is the total heat which is deposited below the photosphere (at $\tau\sim 1$), measured in units of the incident stellar irradiation \citep[adopted from][]{Batygin2011}.
We assume $\epsilon\ll 1$ and $\alpha>0$, consistent with many studies that invoke conversion of a portion of the stellar irradiation into heat deposited deeper inside the atmosphere \citep[][]{GuillotShowman2002,ShowmanGuillot2002,SpiegelBurrows2013}.  
The heat deposition mechanism may have a cut-off at some $\tau_{\rm cut}$ or continue all the way to the center $\tau_c$. In this work we focus on the Ohmic heating mechanism (see Section \ref{sec:physical}), which extends to the planet's deep interior \citep{Batygin2011,HuangCumming2012,WuLithwick2013}. The consequences of a cut-off at $\tau_{\rm cut}<\tau_c$ are discussed in Appendix \ref{sec:other_cases}. 

\begin{figure}[tbh]
	\includegraphics[width=\columnwidth]{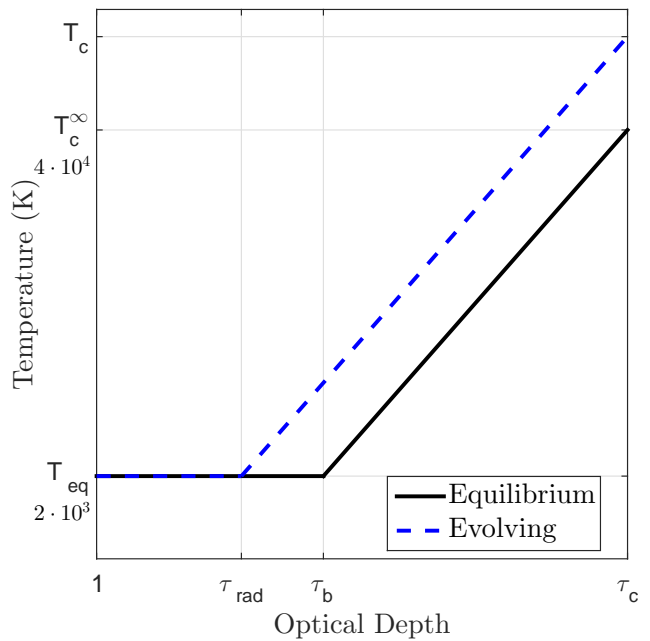}
	\caption{Schematic temperature profile (logarithmic scale) of a hot Jupiter with an energy deposition that extends to its center. The equilibrium state (solid black line) is characterized by an equilibrium central temperature $T_c^\infty$. A hot-Jupiter profile with $T_c>T_c^\infty$ which has not yet reached equilibrium (dashed blue line) is also plotted. The structure of the planet consists of an outer radiative, nearly isothermal region, and a convective interior. Typical values of the temperature are provided.
		\label{fig:scheme_center}}
\end{figure}

The profile in the outer radiative layer follows Equation \eqref{eq:deposit_rad}, which reduces to
\begin{equation}\label{eq:rad_power}
\frac{dU}{d\tau}=\frac{3}{4\pi R^2c}\epsilon L_{\rm eq}\tau^{-\alpha},
\end{equation}
as long as $L_{\rm dep}(\tau)>L_{\rm int}$.
For $\alpha<1$ (see Appendix \ref{sec:other_cases} for other cases), and $\tau\gg 1$ (where the diffusion approximation holds) integration of Equation \eqref{eq:rad_power} from the photosphere inward yields
\begin{equation}\label{eq:u_tau_hat}
U=U_{\rm eq}+\frac{3}{c}\frac{\epsilon L_{\rm eq}}{4\pi R^2}\frac{\tau^{1-\alpha}}{1-\alpha}.
\end{equation}
Therefore, the radiative profile is linear in $\tau^{1-\alpha}$ and some of the results of Section \ref{sec:point} can be reproduced by considering the $[\tau^{1-\alpha},U]$ plane instead of $[\tau,U]$. The convective profile is given by
\begin{equation}\label{eq:power_conv}
U\propto\tau^\beta=(\tau^{1-\alpha})^{\beta/(1-\alpha)},
\end{equation} 
with $\beta/(1-\alpha)$ playing the role of $\beta$ in the analogy with Section \ref{sec:point}. 
We focus on heating profiles which are too flat (decline too gradually with depth) to support a radiative temperature profile $\alpha<1-\beta$ (relevant for Ohmic heating, as discussed in Section \ref{sec:physical}; see Appendix \ref{sec:other_cases} for other values of $\alpha$). In this case, by analogy with Section \ref{sec:point}, convection appears at $\tau_b\sim L_{\rm eq}/L_{\rm dep}(\tau_b)$, or
\begin{equation}\label{eq:power_tau_b}
\tau_b\sim\epsilon^{-1/(1-\alpha)},
\end{equation} 
and radiation energy density $U_{\rm eq}/U=1-\beta/(1-\alpha)$.
From $\tau_b$ the convective region continues to the planet's center $\tau=\tau_c$, reaching a central radiation energy density of
\begin{equation}\label{eq:u_final}
\frac{U_c}{U_{\rm eq}}\sim\left(\frac{\tau_c}{\tau_b}\right)^\beta\sim\left(\tau_c\epsilon^{1/(1-\alpha)}\right)^\beta.
\end{equation}
Thus, the energy deposition dictates an equilibrium central temperature of
\begin{equation}\label{eq:t_final}
\frac{T_c^\infty}{T_{\rm eq}}\sim\left(\tau_c\epsilon^{1/(1-\alpha)}\right)^{\beta/4},
\end{equation}
and according to Section \ref{subsec:radius}, a final planet radius of $\Delta R^\infty\sim k_{\rm B}T_c^\infty/m_pg$, at which evolutionary cooling stops. A schematic profile of a planet in this equilibrium state is given in Figure \ref{fig:scheme_center}.

For a planet with a central temperature $T_c>T_c^\infty$, it is easy to see from Equations \eqref{eq:irradiated}, \eqref{eq:u_final}, and from Figure \ref{fig:scheme_center} that $\tau_{\rm rad}<\tau_b$ and that $L_{\rm dep}(\tau_{\rm rad})<L_{\rm int}$. Therefore, the radiative-convective transition and the internal luminosity in this case are unaffected by the energy deposition, and are determined as in Section \ref{subsec:irradiation}. 

In summary, deep heat deposition, which extends to the planet's interior, does not slow down the cooling rate of the planet, but rather imposes a high equilibrium central temperature (and radius), at which the planet stops evolving entirely \citep[see also][]{Batygin2011}. The cooling history of the planet until it reaches this equilibrium state is unaffected by the deposited energy (see Appendix \ref{sec:other_cases} for a more general discussion). Using Equation \eqref{eq:power_tau_b}, we find that the planet cools down, and $\tau_{\rm rad}$ increases (see Section \ref{subsec:irradiation}), until 
\begin{equation}\label{eq:crit_tau_rad}
\epsilon\tau_{\rm rad}^{1-\alpha}\gtrsim 1,
\end{equation}
or, equivalently, $L_{\rm dep}(\tau_{\rm rad})\gtrsim L_{\rm int}$, meaning that the deposited heat in the convective region exceeds the internal luminosity \citepalias[see][for analogy with the point-source deposition]{GS2015}. Similar results are found by \citet{WuLithwick2013}. Since the radiative-convective boundary of $\sim 1\textrm{ Gyr}$ old irradiated planets with a typical equilibrium temperature of $T_{\rm eq}\approx 2\cdot 10^3\textrm{ K}$ lies at an optical depth of $\tau_{\rm rad}\approx 10^5$ in the absence of power deposition \citep[see, e.g.,][]{ArrasBildsten2006,Batygin2011,GS2015}, the critical efficiency required to inflate observed planets can be roughly estimated as $\epsilon\gtrsim 10^{-5(1-\alpha)}$.

\section{Application to Ohmic Heating}
\label{sec:physical}

We now apply the results of Section \ref{sec:power_law} to the Ohmic heating mechanism \citep{BatyginStevenson2010,Perna2010,Perna2012,Batygin2011,HuangCumming2012,RauscherMenou2013,WuLithwick2013,RogersShowman2014}.

\subsection{Atmospheric Winds and Induced Currents}
\label{subsec:winds_currents}

\begin{figure}[tbh]
\includegraphics[width=\columnwidth]{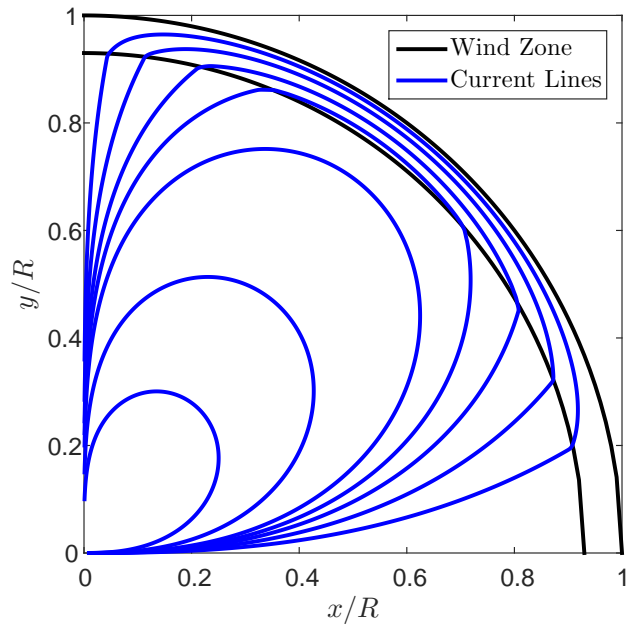}
\caption{Schematic current surface density $J$ field line representation. $J$ is given by the solution to Equation \eqref{eq:ohm_law}, under the assumption of a dipolar magnetic field, and a constant velocity zonal wind, confined to a shallow wind zone \citep[see][for a detailed solution]{WuLithwick2013}. The width of the wind zone is exaggerated. Due to the $l=2$, $m=0$ symmetry, only the first quadrant of a meridional plane is displayed. Below the wind zone the radial and tangential components are comparable, since $\nabla\cdot(\sigma\nabla\Phi)=0$, while at the surface the radial component vanishes.  
\label{fig:lines}}
\end{figure}

The electric current surface density $\vec{J}$ is related to the planet's magnetic field $\vec{B}$ and to the atmospheric wind velocity $\vec{v}$ through Ohm's law 
\begin{equation}\label{eq:ohm_law}
\vec{J}=\sigma\left(-\nabla\Phi+\frac{\vec{v}}{c}\times \vec{B}\right),
\end{equation}
with $\sigma$ denoting the conductivity, and $\Phi$ the induced electric potential. For a simple approximate model, we follow \citet{WuLithwick2013} and assume a constant-velocity wind, confined to a shallow wind-zone with thickness determined by the isothermal atmosphere scale height $H=k_{\rm B}T_{\rm eq}/m_pg\approx 10^8\textrm{ cm}\ll R$. \citet{WuLithwick2013} denote the wind-zone depth by an arbitrary $z_{\rm wind}$ (for which they choose fiducial values $\sim 10^8\textrm{ cm}$), but as we show below, $z_{\rm wind}\sim H$ \citep[see also][]{Batygin2011}. The potential $\Phi$ and the current surface density $\vec{J}$ are found by applying the continuity equation in steady state $\nabla\cdot \vec{J}=0$ to Equation \eqref{eq:ohm_law}. The solution to this model is described in detail by \citet{WuLithwick2013}, and is characterized by a current density magnitude of
\begin{equation}\label{eq:j_size}
J(r)\sim\begin{cases}
J_0  & R-r<H \\
J_0\frac{H}{R}\left(\frac{r}{R}\right)^{l-1}\sim J_0\frac{H}{R} & R-r>H 
\end{cases},
\end{equation}
with 
\begin{equation}\label{eq:j0}
J_0\equiv\sigma\frac{v}{c}B
\end{equation}
evaluated in the wind zone and with a discontinuous drop of order $H/R$ over the edge of the wind zone. This current drop is the result of the outer boundary condition (the radial component of the current vanishes at $r=R$) and the solution to Equation \eqref{eq:ohm_law} below the wind zone $\nabla\cdot(\sigma\nabla\Phi)=0$, with an $l=2$ symmetry imposed by the $\vec{v}\times \vec{B}$ term \citep[assuming a dipolar magnetic field and zonal winds; see, e.g.,][]{BatyginStevenson2010,WuLithwick2013}, though this result can be generalized to other geometries. In Figure \ref{fig:lines} we present a schematic plot of the current field lines, which must form closed loops (since $\nabla\cdot \vec{J}=0$). The $H/R$ current drop is intuitively understood due to the folding of the current lines inside the wind zone, in order to eliminate their radial component at the surface (note that the wind zone depth is exaggerated in Figure \ref{fig:lines}). Alternatively, the drop can be understood by noting that the condition $\nabla\cdot\vec{J}=0$ leads to $J_r/H\sim J_{\theta}/R$ (with $J_\theta$, $J_r$ denoting the tangential and radial components of the current) in the wind zone and that $J_r$ transitions continuously below the wind zone, where it is comparable in magnitude to the tangential component. Since $J(r)\propto(r/R)^{l-1}$ below the wind zone, the current surface density is roughly constant there (while the pressure varies by orders of magnitude) as long as the radius is not much smaller than $R$. The decrease of the current, and therefore the Ohmic dissipation, at $r\ll R$ is irrelevant for the planet's inflation, because the density, and therefore the temperature, approach their maximal (central) values at $r\sim R/2$ in a polytropic profile \citep[e.g.,][]{Peebles64}. We note that our two layer model is very similar to the three layer analytical model of \citet{WuLithwick2013}. Instead of a discontinuous jump in the conductivity with depth (between their middle and inner layers), we use a smooth power-law, as described in Section \ref{subsec:ohmic_power_law}, which better represents numerical conductivity profiles calculated in previous studies \citep{Batygin2011,HuangCumming2012}, including \citet{WuLithwick2013} itself.

The acceleration of a fluid element due to the Lorentz force is given by
\begin{equation}\label{eq:lorentz}
\vec{f}=\frac{1}{\rho}\vec{J}\times\frac{\vec{B}}{c}.
\end{equation}
By combining Equations \eqref{eq:ohm_law} and \eqref{eq:lorentz} we find the magnetic drag deceleration
\begin{equation}\label{eq:magnetic_drag}
\vec{f}=-\frac{B^2}{\rho c^2}\sigma\vec{v}.
\end{equation}
Following \citet{BatyginStevenson2010} and \citet{Laughlin2011}, we estimate the magnetic field of the planet using the Elsasser number criterion \citep[see][for an alternative]{Christensen2009}, which arises from a balance between the Lorentz and Coriolis forces at the core
\begin{equation}\label{eq:elsasser}
\frac{B^2}{\rho_c c^2}=\frac{\Omega}{\sigma_c},
\end{equation}
where the Lorentz force is given by Equation \eqref{eq:magnetic_drag}, but with the conductivity $\sigma_c$ and density $\rho_c$ of the planet's interior instead of the atmosphere. $\Omega\approx 10^{-5}\textrm{ s}^{-1}$ denotes the rotation frequency of the planet, which is equal to its orbital frequency since close-in planets are tidally locked \citep[see, e.g.][]{Guillot96}.

The atmospheric flow velocity $v$ should be determined by global circulation models, coupled with magnetic fields \citep[see, e.g.][]{RogersKomacek2014, RogersShowman2014}.
For example, one effect which is taken into account in these simulations, but neglected in our following analysis, is the induced (by currents) magnetic field, which should be added to the planet's dipolar field. 
Nonetheless, we make a rough order of magnitude estimate here, following \citet{ShowmanGuillot2002}. The winds are driven by a horizontal temperature difference $\Delta T\lesssim T_{\rm eq}$ between the day and night sides of the tidally locked planet, leading to a forcing acceleration of $(c_s^2/R)(\Delta T/T_{\rm eq})$, with $c_s\sim(k_{\rm B}T_{\rm eq}/m_p)^{1/2}$ denoting the speed of sound in the atmosphere. This forcing is balanced by both the Coriolis force \citep[the classical thermal wind equation; see, e.g.][]{ShowmanGuillot2002,Showman2010} and the magnetic drag \citep{Batygin2011,Menou2012}
\begin{equation}\label{eq:wind_vel}
\frac{c_s^2}{R}\frac{\Delta T}{T_{\rm eq}}=\Omega v\left(1+\frac{\sigma}{\sigma_c}\frac{\rho_c}{\rho}\right),
\end{equation}
where the magnetic drag is given by Equations \eqref{eq:magnetic_drag} and \eqref{eq:elsasser}. Unlike \citet{Menou2012}, we neglect the nonlinear advective term $v\nabla v\sim v^2/R$ with respect to the Coriolis force, since the Rossby number is $\textrm{Ro}\sim v/(\Omega R)<1$, as evident from Equation \eqref{eq:wind_vel} \citep[see also][]{ShowmanGuillot2002,Showman2010}. Note that ${\rm Ro}\sim 1$ only when the magnetic drag is negligible, and the temperature difference is maximal $\Delta T\approx T_{\rm eq}$, since (by coincidence) the sound speed and rotation velocity are similar $c_s\sim\Omega R\sim 10^5\textrm{ cm s}^{-1}$. Therefore, a low Rossby number approximation is more adequate for the general case ($\Delta T\le T_{\rm eq}$, and with magnetic drag included). A more careful analysis takes into account the dependence of the Rossby number (the ratio between the advective and Coriolis terms) on the latitude $\phi$: $\textrm{Ro}=v/(2\Omega R\sin\phi)$. Evidently, our low Rossby number approximation is valid except for a narrow ring around the equator $\phi\to 0$, therefore providing a reasonable estimate for the average atmospheric behavior. An alternative analysis, relevant close to the equator, and in which the advective term is dominant, is presented in Appendix \ref{sec:equator}. As seen in Appendix \ref{sec:equator}, both methods lead to qualitatively similar results \citep[see also][]{Menou2012}.

Equation \eqref{eq:wind_vel} indicates the existence of two regimes: an unmagnetized regime, where the forcing is balanced by the Coriolis force, and a magnetized regime, in which the magnetic drag is the balancing force. The transition between the regimes is when $\sigma/\sigma_c\sim\rho/\rho_c$. Due to the sharp increase of the conductivity with temperature (see Section \ref{subsec:ohmic_power_law}), the magnetized regime corresponds to high equilibrium temperatures. Using the conductivities calculated by \citet{HuangCumming2012}, we estimate the transition equilibrium temperature at $T_{\rm eq}\approx 1500\textrm{ K}$, inside the observationally relevant range \citep[see][for similar results]{Menou2012}. 

The temperature difference $\Delta T$ is determined, in the simplest analysis, by the ratio of advective to radiative timescales \citep{ShowmanGuillot2002,Menou2012}, with higher-order effects taken into account by more comprehensive treatments \citep{PBS2013,KomacekShowman2015}. Explicitly, we write a diffusion equation for the change of $\delta T\equiv T-T_{\rm eq}$ with time and optical depth $\tau$, taking into account the atmospheric thermal inertia (and therefore the radiative timescale)
\begin{equation}\label{eq:diffusion}
\frac{\partial \delta T}{\partial t}=\kappa m_p\frac{\sigma_{\rm SB} T_{\rm eq}^3}{k_{\rm B}}\frac{\partial^2\delta T}{\partial\tau^2}.
\end{equation}
Note that by writing the diffusion equation using the optical depth, instead of the spatial coordinate, we take into account the variation of the diffusion coefficient with depth.
We assume a solution of the form $\delta T=\Delta Te^{i\omega(\tau)t-k(\tau)\tau}$, where $\omega(\tau)$ and $k(\tau)$ are power laws, and the periodic temporal dependence is determined by the advective timescale between the day and night sides $\omega^{-1}\equiv R/v$. We solve equations \eqref{eq:wind_vel} and \eqref{eq:diffusion} together, and find the decay of the day-night temperature difference $\Delta T$ with optical depth, up to a logarithmic factor.
\begin{equation}\label{eq:delta_t}
\frac{\Delta T}{T_{\rm eq}}=\frac{R^2\sigma_{\rm SB}T_{\rm eq}^4\Omega}{c_s^4}\frac{\kappa}{\tau^2}\left(1+\frac{\sigma}{\sigma_c}\frac{\rho_c}{\rho}\right).
\end{equation} 
Using Equation \eqref{eq:delta_t} and atmospheric opacities from \citet{Allard2001} and \citet{Freedman2008}, we find that the condition $\Delta T\sim T_{\rm eq}$ is satisfied near the photosphere ($\tau\sim 1$) for our fiducial $T_{\rm eq}\approx 2\cdot 10^3\textrm{ K}$ \citep[see][for the same conclusion]{ShowmanGuillot2002, Menou2012}. This result explains observed day-night temperature differences \citep[see, e.g.,][]{Knutson2009}, which are smaller than $T_{\rm eq}$, but only by an order of unity factor. \citet{ShowmanGuillot2002} and \citet{Menou2012} obtain the same result using a simpler prescription for the radiative timescale, which is different by a factor of $\tau$ and valid only for $\tau=1$, coinciding with Equation \eqref{eq:delta_t} in this case. Our factor of $\tau^2$, as well as a more intuitive approach to deriving Equation \eqref{eq:delta_t}, is also obtained by explicitly writing the radiative timescale at an optical depth $\tau$
\begin{equation}
t_{\rm rad}(\tau)=\frac{E}{L}\sim\frac{\frac{R^2\tau}{\kappa m_p}k_{\rm B}T_{\rm eq}}{R^2\sigma_{\rm SB} T_{\rm eq}^4/\tau}=\frac{c_s^2\tau^2}{\kappa\sigma_{\rm SB} T_{\rm eq}^4},
\end{equation}
where one factor of $\tau$ \citep[which is taken into account by][]{ShowmanGuillot2002} is due to the mass, and therefore the energy, of a layer of thickness $\tau$, while a second factor of $\tau$ is due to the luminosity through this optical depth $L\sim R^2\sigma_{\rm SB}T_{\rm eq}^4/\tau$ \citep[see also][for numerical radiative timescale calculations]{Iro2005,Showman2008}.
Interestingly, Equation \eqref{eq:delta_t} indicates that for low temperatures (and therefore, in the unmagnetized regime) the relative day-night temperature difference at the photosphere ($\tau\sim 1$) falls rapidly with decreasing temperatures $\Delta T/T_{\rm eq}\propto T_{\rm eq}^5$, where we neglect the weak dependence of the photospheric opacity on the temperature, and substitute $\Omega\propto T_{\rm eq}^3$ (though at large enough separations the planets may not be tidally locked). Consequently, we predict that warm Jupiters, with $T_{\rm eq}\lesssim 10^3\textrm{ K}$ will have very small day-night temperature differences, compared to hot Jupiters, with $T_{\rm eq}\gtrsim 10^3\textrm{ K}$, which are in the saturated regime $\Delta T\sim T_{\rm eq}$ of Equation \eqref{eq:delta_t}. This prediction, which is robust to the Rossby number regime, as seen by Equation \eqref{eq:delta_t_high_ro}, can be tested against observations \citep[e.g.,][]{Kammer2015,Wong2015}.
Using Equations \eqref{eq:wind_vel} and \eqref{eq:delta_t}, we find the decay of the velocity with depth $v\propto\Delta T\propto\kappa/\tau^2\propto P^{-5/2}$, with $P$ denoting the pressure, and utilizing the relation $P/g\sim\tau/\kappa$ (we neglect the mild change of $\sigma/\rho$ with depth in the atmosphere for the magnetized case). 

The velocity drops as a power law in pressure, which increases exponentially with a scale height $H$, allowing us to replace the discontinuous drop in the current surface density in the \citet{WuLithwick2013} model with a continuous drop from an outer $J=J_0$, given by Equation \eqref{eq:j0}, to a roughly constant $J=J_0(H/R)$ in the interior. 
The Ohmic dissipation per unit mass is given by
\begin{equation}\label{eq:ohmic}
\frac{dL}{dm}=\frac{J^2}{\rho\sigma},
\end{equation}
implying a drop of order $(H/R)^2$ in the dissipated power.
Since \citet{WuLithwick2013} choose fiducial values $z_{\rm wind}\sim 10^8\textrm{ cm}\sim H$, we predict a similar decrease in the Ohmic dissipation, without introducing the arbitrary $z_{\rm wind}$ parameter.
In contrast to \citet{WuLithwick2013} and to this work, the model of \citet{Batygin2011} does not predict a drop in the current at the edge of the wind zone. This results in their overestimation of the deposition at depth, though it is partially balanced by their slightly steeper heating profile (see Section \ref{subsec:ohmic_power_law}).     

\subsection{Ohmic Deposition as a Power Law}
\label{subsec:ohmic_power_law}

Since the current density outside the wind zone is roughly constant, using Equation \eqref{eq:ohmic}, the Ohmic dissipation power-law is determined by $dL/dm\propto 1/(\rho\sigma)$.
The electric conductivity in the outer layers of the planet is determined by the ionization level of alkali metals, with Potassium dominating the results, due to its low ionization energy \citep{HuangCumming2012}. \citet{HuangCumming2012} calculated the Potassium ionization level using the Saha equation, and arrived at the scaling of the conductivity with pressure and temperature. For a simple power-law estimate, we evaluate their scaling (their Equation A3) at the radiative convective boundary as $\sigma\propto T^{7.5}\rho^{-0.5}$ (the exponential dependence on the temperature is approximated as a power law with index $={\rm d}\ln\sigma(T)/{\rm d}\ln T$, evaluated at the radiative-convective boundary), and obtain the relation $\sigma\propto P^{0.8}$, where we used $P\propto T^{n+1}$, with $n\approx 5$ from the equation of state of \citet{Saumon95}. This polytropic equation of state is a good approximation in the relevant temperature range, as seen, for example, in Figure 2 of \citet{Batygin2011}, who also incorporate the \citet{Saumon95} equation of state. We note that in \citetalias{GS2015} we used a different $n\approx 2$, relevant for lower radiative-convective boundary temperatures that characterize less irradiated and unheated planets (the results of \citetalias{GS2015} depend weakly on $n$). Although our modeling of the conductivity as a power-law is an ad-hoc simplifying approximation, realistic conductivities exhibit a (roughly) power-law dependence on the pressure level inside the planet \citep[see, e.g., Figure 3 of][]{Batygin2011}.  
Using Equation \eqref{eq:ohmic} and $\rho\propto P^{n/(n+1)}$, we find that the specific Ohmic dissipation scales as $dL/dm\propto P^{-1.6}$. This scaling is in between the somewhat steeper scaling of \citet{Batygin2011} and the flatter scaling of \citet{WuLithwick2013}. The accumulated luminosity therefore scales as $L\propto P^{-0.6}$ (since pressure is linear in mass at the outer edge of the planet).

In order to find $\alpha$ we relate the pressure to the optical depth $P\propto T^{n+1}\propto\tau^{\beta(n+1)/4}$. We estimate $\beta\approx 0.35$ using Equation \eqref{eq:beta} and the opacities of \citet{Freedman2008}, in the vicinity of the radiative-convective boundary \citepalias[see also][for similar results]{GS2015}. The resulting power law is $\alpha\approx 0.3$, which is in the $\alpha<1-\beta$ regime (see Section \ref{sec:power_law} and Appendix \ref{sec:other_cases}).

\subsection{Ohmic Dissipation Efficiency}
\label{subsec:epsilon}

\begin{figure}[tbh]
\includegraphics[width=\columnwidth]{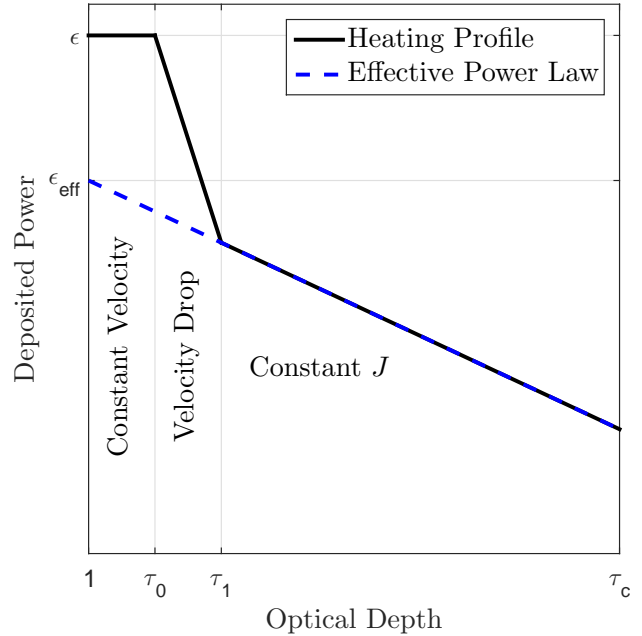}
\caption{Schematic Ohmic heating profile (logarithmic scale) of a hot Jupiter.
The profile (solid black line) is given by the integrated power, deposited below optical depth $\tau$, in units of the incident stellar irradiation. The profile is characterized by three distinct regions: a constant velocity wind zone up to $\tau_0\sim 1$, a velocity drop from $\tau_0$ to $\tau_1$, and a constant current surface density $J$ region in the interior. An effective power law heating profile which defines $\epsilon_{\rm eff}$ (dashed blue line) is also plotted.
\label{fig:eps_eff}}
\end{figure}

In this section we estimate the efficiency $\epsilon$ in the Ohmic dissipation scenario, following the arguments of \citet{Batygin2011} and \citet{HuangCumming2012}. As we discuss below (see Figure \ref{fig:eps_eff}), this efficiency is dominated by the dissipation in the wind zone, and is higher than the effective efficiency $\epsilon_{\rm eff}$ (which is related to $\epsilon$ below) used to evaluate the deep energy deposition.
Using Equations \eqref{eq:ohm_law} and \eqref{eq:ohmic}, the efficiency, which is defined as the dissipated energy rate in units of the stellar irradiation, is given by
\begin{equation}\label{eq:epsilon_sigma}
\epsilon=\frac{\left(v/c\right)^2\sigma B^2H}{\sigma_{\rm SB}T_{\rm eq}^4}=\frac{\left(\sigma/\sigma_c\right)\rho_cv^2H\Omega}{\sigma_{\rm SB}T_{\rm eq}^4},
\end{equation}
where we have eliminated the magnetic field using Equation \eqref{eq:elsasser}.
This result can also be understood by dividing the kinetic energy by the magnetic drag's stopping time $(\rho c^2/B^2)\sigma^{-1}$, which is obtained from Equation \eqref{eq:magnetic_drag}.

It is instructive to consider the variation of the dominant term in the efficiency $\sigma v^2$ with conductivity, while assuming all other parameters constant. From Equation \eqref{eq:wind_vel} we find
\begin{equation}\label{eq:epsilon_power_law_sigma}
\sigma v^2\propto\begin{cases}
\sigma & \sigma<\sigma_m \\[0.5ex]
\sigma^{-1} & \sigma>\sigma_m
\end{cases},
\end{equation}
with $\sigma_m\sim\sigma_c(\rho/\rho_c)\sim 10^9\textrm{ s}^{-1}$ denoting the transition between the magnetized and unmagnetized regimes \citep[see][for similar results]{Menou2012}. The maximal efficiency, obtained at $\sigma_m$ is
\begin{equation}\label{eq:epsilon_max}
\epsilon_{\rm max}=\frac{\rho v^2H\Omega}{\sigma_{\rm SB}T_{\rm eq}^4}=\frac{\rho c_s^4H/\Omega}{4R^2\sigma_{\rm SB}T_{\rm eq}^4}\approx\frac{1}{4\tau_0}\approx 0.3,
\end{equation}
with $\tau_0\sim 1$ denoting the optical depth where the day-night temperature difference falls below order unity, which is obtained by setting $\Delta T/T_{\rm eq}=1$ in Equation \eqref{eq:delta_t}, and with the density $\rho$ given by the condition $\tau_0\sim\kappa\rho H$.
 
This maximal efficiency is similar to \citet{Menou2012}, who neglected the Coriolis force. \citet{Menou2012} also decoupled the magnetic field strength from the rotation of the planet (and therefore the equilibrium temperature). In our approach, on the other hand, the magnetic field is related to $\Omega\propto T_{\rm eq}^3$ through the Elssaser number condition, as described above. By taking this relation into account, incorporating the scaling of conductivity $\sigma$ with temperature from Section \ref{subsec:ohmic_power_law}, and considering the dependence of all other variables: $H$, $\rho$, and $c_s$ on the temperature, Equations \eqref{eq:wind_vel} and \eqref{eq:epsilon_sigma} yield
\begin{equation}\label{eq:epsilon_power_law}
\epsilon\propto\begin{cases}
T_{\rm eq}^3 & T_{\rm eq}<T_m \\[1.5ex]
T_{\rm eq}^{-6} & T_{\rm eq}>T_m
\end{cases},
\end{equation}
with $T_m\approx 1500\textrm{ K}$ denoting the transition between the magnetized and unmagnetized regimes. Equation \eqref{eq:epsilon_power_law} has the same qualitative behavior as the more illustrative Equation \eqref{eq:epsilon_power_law_sigma}, which demonstrates the dependence of the efficiency on the conductivity $\sigma$ (which is the dominant factor, due to its sharp increase with temperature).

The efficiency $\epsilon$ above denotes the total dissipated energy rate in units of the stellar irradiation. However, the formalism of Section \ref{sec:power_law} assumes a single power-law deposition profile, while the actual heating function is a broken power-law with three segments, as shown in Figure \ref{fig:eps_eff}. Nevertheless, we can use  Section \ref{sec:power_law}, by replacing  $\epsilon$ with $\epsilon_{\rm eff}$. The first segment, characterized by a flat deposition due to the increase of conductivity with depth (since $L\propto J^2/\sigma$, $J\propto\sigma v$ in the wind zone, and $v$ is constant for $\tau<\tau_0$), extends to $\tau_0\sim 1$. Beyond $\tau_0$, the velocity drops as $v\propto P^{-5/2}$, so $J\propto P^{-1.7}\propto\tau^{-0.9}$ until $\tau_1=\tau_0(H/R)^{-1.1}$, beyond which $J$ is roughly constant. Combining these factors, we find $\epsilon_{\rm eff}\approx\epsilon\tau_0^\alpha(H/R)^2\approx 2\cdot 10^{-3} \epsilon$. 

\subsection{Implications for Planet Inflation}
\label{subsec:inflation}

We estimate an effective critical heating efficiency of $\epsilon_{\rm eff}\approx 10^{-4}$, required to inflate observed ($\approx 3$ Gyr old with $T_{\rm eq}\approx 2\cdot 10^3\textrm{ K}$) hot Jupiters, by substituting $\alpha\approx 0.3$ and $\tau_{\rm rad}\approx 10^5$ in Equation \eqref{eq:crit_tau_rad}. By taking into account the translation between $\epsilon_{\rm eff}$ and $\epsilon$ in the Ohmic scenario (see Section \ref{subsec:epsilon}), our estimate for the actual critical efficiency is $\epsilon\approx 5\%$, consistent with \citet{Batygin2011} and \citet{WuLithwick2013}. However, our results agree with \citet{Batygin2011} only because the absence of an electric-current drop in their model is balanced by a steeper heating profile.  
 
More concretely, Equation \eqref{eq:t_final} predicts an equilibrium central temperature of
\begin{equation}\label{eq:t_final_ohm}
\frac{T_c^\infty}{T_{\rm eq}}\sim\left[\tau_m\left(\frac{T_{\rm eq}}{T_m}\right)^b\epsilon_{\rm eff}^{1/(1-\alpha)}\right]^{\beta/(4-\beta b)},
\end{equation} 
where we normalize the relation $\tau_c/\tau_m=(T_c/T_m)^b$ \citepalias[due to the opacity scaling with the temperature $\tau_c\propto\kappa_c\propto T_c^b$; see][and Section \ref{subsec:irradiation}]{GS2015} to the temperature of the magnetic transition $T_m$ and to the corresponding optical depth $\tau_m\approx 10^{10}$. As discussed in Section \ref{subsec:radius}, this temperature implies an equilibrium radius of
\begin{equation}\label{eq:eq_radius}
\Delta R^\infty\approx 0.3 R_J\left(\frac{\epsilon}{5\%}\right)^{0.3}\left(\frac{T_{\rm eq}}{1500\textrm{ K}}\right)^3,
\end{equation}
where we substitute $\beta=0.35$, $b=7$ from \citetalias{GS2015}, and with the transition between $\epsilon$ and $\epsilon_{\rm eff}$ accounted for.

\begin{figure}[tbh]
\includegraphics[width=\columnwidth]{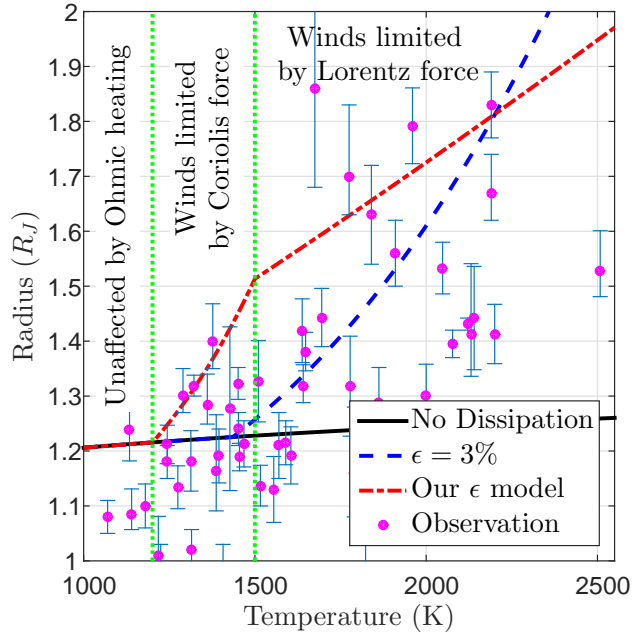}
\caption{Radius inflation of 3 Gyr old Jupiter-mass planets due to Ohmic dissipation, as a function of their equilibrium temperature. The radius is given by the maximum of the equilibrium radius imposed by the heat deposition, according to Equation \eqref{eq:eq_radius}, and the radius which cooling under the influence of stellar irradiation predicts (solid black line), according to Equation \eqref{eq:rad_cooling}. The constant efficiency curve (dashed blue line) is given by $\epsilon=3\%$, while the variable efficiency curve (dot-dashed red line) is given by Equation \eqref{eq:epsilon_power_law} with $\epsilon_{\rm max}=0.3$, and $T_m=1500\textrm{ K}$. The magenta points, taken from the exoplanet.eu database, correspond to observed 0.5-2.0 $M_J$ planets.
\label{fig:power_law}}
\end{figure}

The planet's radius as a function of time, before the planet reaches equilibrium is given in \citetalias{GS2015}:
\begin{equation}\label{eq:rad_cooling}
\Delta R(t)\approx 0.2 R_J\left(\frac{T_{\rm eq}}{1500\textrm{ K}}\right)^{0.25}\left(\frac{t}{5\textrm{ Gyr}}\right)^{-0.25}.
\end{equation}
Comparison of Equations \eqref{eq:eq_radius} and \eqref{eq:rad_cooling} shows that $\epsilon\gtrsim 5\%$ can explain the majority of observed radius discrepancies. In Figure \ref{fig:power_law} we present an estimate of the radii of hot Jupiters which are Ohmically heated with a constant efficiency of 3\%. Our analytical model roughly reproduces the numerical results of \citet{WuLithwick2013} in this scenario, with differences explained by the flatter heating profile and the absence of coupling between the wind zone depth and the temperature in \citet{WuLithwick2013}. 
However, the ad-hoc assumption of some constant efficiency is inadequate for the Ohmic dissipation mechanism, as explained in Section \ref{subsec:epsilon} and \citet{Menou2012}. For this reason, we also present in Figure \ref{fig:power_law} a more comprehensive, variable efficiency model, according to Equations \eqref{eq:epsilon_max} and \eqref{eq:epsilon_power_law}.
This variable $\epsilon$ model predicts that Ohmic dissipation inflates planets with equilibrium temperatures $\gtrsim 1500\textrm{ K}$ to a radius $\gtrsim 1.5 R_J$.
These results are in agreement with most of the observations, with a few extremely bloated planets remaining unexplained by Ohmic dissipation \citep[see also][]{WuLithwick2013}.  
Specifically, the variable efficiency may explain the excess of observed radius anomalies at $T_{\rm eq}\gtrsim 1500\textrm{ K}$ \citep[see also][]{DemorySeager2011,Laughlin2011,MillerFortney2011,Schneider2011}. 

In this work we limited ourselves, for simplicity, to roughly Jupiter-mass planets. Gas giants with a different mass, but still in the regime $M\sim M_J$, are located near the inversion of the zero temperature radius-mass relation \citepalias[see, e.g.,][]{GS2015}, and their density can therefore be modeled approximately as $\rho\propto M/R^3\propto M$. By inserting this relation in Equations \eqref{eq:delta_r_t} and \eqref{eq:t_final} we estimate $\Delta R^\infty\propto M^{(a+1)\beta/(4-\beta b)-1}=M^{-0.6}$, with $a=0.5$ \citepalias{GS2015}. This result is similar to the fit of \citet{WuLithwick2013}, and can explain the large radii (up to $\approx 1.8 R_J$) of many low-mass ($\approx 0.4 M_J$) inflated planets (with the small decrease in the zero-temperature radius taken into account). However, some of the most inflated hot Jupiters depicted in Figure \ref{fig:power_law} have a large mass $M\gtrsim 0.8M_J$ \citep[see, e.g.,][]{Baraffe2014}, and therefore exceed the predictions of our model. 

\subsection{Comparison with Previous Works}
\label{subsec:comparison}

In this section we summarize the main qualitative differences between this work and previous studies of the Ohmic heating mechanism.

The total dissipated power, given by the efficiency $\epsilon$, depends on the speed of atmospheric flows. \citet{Batygin2011} and \citet{WuLithwick2013} assume fiducial speeds of $\sim 1\textrm{ km s}^{-1}$, which result in $\epsilon\approx 3\%$. However, \citet{WuLithwick2013} advocate for a discontinuous drop of order $(z_{\rm wind}/R)^2\sim 10^{-3}$ in the dissipated power over the wind zone edge, which is absent in the model of \citet{Batygin2011}. 

In this work we studied the decay of atmospheric winds with depth, by comparing the advective and radiative timescales. In contrast to previous studies \citep{ShowmanGuillot2002, HuangCumming2012, Menou2012} our treatment is valid for optical depths above unity. We found that the winds decay as a power law with pressure, allowing us to replace the discontinuity of \citet{WuLithwick2013} with a continuous drop, and to relate the wind zone depth $z_{\rm wind}$ \citep[which is an arbitrary parameter in][]{WuLithwick2013} to the atmospheric scale height.  

In contrast to \citet{WuLithwick2013}, we do not choose a constant fiducial wind speed, and therefore a constant efficiency, but rather calculate the wind speed by comparing the thermal forcing (due to the day-night temperature difference) to the Coriolis force (relevant for $T_{\rm eq}\lesssim 1500\textrm{ K}$) and to the magnetic drag (relevant for $T_{\rm eq}\gtrsim 1500\textrm{ K}$). This analysis leads us to replace the constant $\epsilon$ with an efficiency which rises to a maximum of $\epsilon\approx 0.3$ at $T_{\rm eq}\approx 1500\textrm{ K}$, and then drops at higher equilibrium temperatures due to the reduction of wind speeds by the magnetic drag. Our variable $\epsilon$ model is similar to \citet{Menou2012}, with two main differences: we consider the balance between the thermal forcing and the Coriolis force instead of the nonlinear advective term $v\nabla v$ (our low Rossby number approximation is more appropriate in this case, see Section \ref{subsec:winds_currents}), and we do not treat the planet's magnetic field as a free parameter, but rather couple it to the planet's rotation rate, and therefore to its equilibrium temperature. Despite these differences, our model qualitatively reproduces the results of \citet{Menou2012}, as seen by comparing our Equation \eqref{eq:epsilon_power_law} to their Figure 4. 

Another new ingredient in our work is the analytic translation of a given heat dissipation efficiency to an inflated planet radius. Most previous studies calculate the Ohmically heated planet evolution using stellar evolution codes, with the exception of \citet{HuangCumming2012}, who numerically integrate a simplified model based on \citet{ArrasBildsten2006}. In this work, however, we calculate the planet's evolution using a generalization of a simple analytic theory, derived in \citetalias{GS2015}. This analytical approach provides a broader understanding of the scaling laws governing hot-Jupiter inflation. 

One interesting example of an insight gained by our analytical model is the qualitative shape of the $R(T_{\rm eq})$ curve plotted in Figure \ref{fig:power_law}. Specifically, how come the inflation increases with temperature, while the heating efficiency $\epsilon(T_{\rm eq})$ drops due to the magnetic drag at high temperatures, as explained above? The answer is obtained by considering Equation \eqref{eq:t_final_ohm}, which indicates that the inflation is determined by two competing factors. While the efficiency decreases, the increasing ${\rm H}^-$ opacity pushes the radiative-convective boundary (in the final equilibrium state) to lower pressures (since $P/g\sim\tau/\kappa$), raising the temperatures of the inner adiabat (this effect is represented by the $T_{\rm eq}^b$ term in the equation).
	
Quantitatively, due to its high maximal efficiency $\epsilon_{\rm max}\approx 0.3$, our model predicts somewhat more inflated planets in the range $1500\textrm{ K}\leq T_{\rm eq}\leq 2000\textrm{ K}$, compared with the constant $\epsilon=3\%$ model of \citet{WuLithwick2013}, as seen by comparing our Figure \ref{fig:power_law} and their Figure 5. However, the differences between the two models are modest ($\approx 0.1 R_J$), due to the relatively weak dependence of the inflated radius on the efficiency $\Delta R^\infty\propto\epsilon^{0.3}$, the sharp drop of the efficiency at $T_{\rm eq}\gtrsim 1500\textrm{ K}$, evident from Equation \eqref{eq:epsilon_power_law}, and the somewhat flatter heating profile of \citet{WuLithwick2013}. The model of \citet{Batygin2011}, on the other hand, predicts higher inflations, due to the absence of a drop in their heating profile (see Figure \ref{fig:eps_eff}). Nonetheless, by comparing our Figure \ref{fig:power_law} to their Figures 6-8, we find that the differences are partially compensated by the lower efficiencies $\epsilon\leq 5\%$ and the somewhat steeper heating profile of \citet{Batygin2011}, and amount to $\approx 0.3R_J$ for $1M_J$ planets.

\section{Planet Re-inflation}
\label{sec:reinflation}

In the previous sections we assumed that planets cool and contract from high temperatures (and therefore large radii) under the influence of both stellar irradiation and additional power deposition (e.g. Ohmic dissipation). However, another possible scenario (due to migration on long timescales, for example) involves dissipation mechanisms that come into play only once the planet has already cooled and contracted to a relatively small radius \citep[see also][and references within]{Batygin2011,WuLithwick2013}.

\begin{figure}[tbh]
\includegraphics[width=\columnwidth]{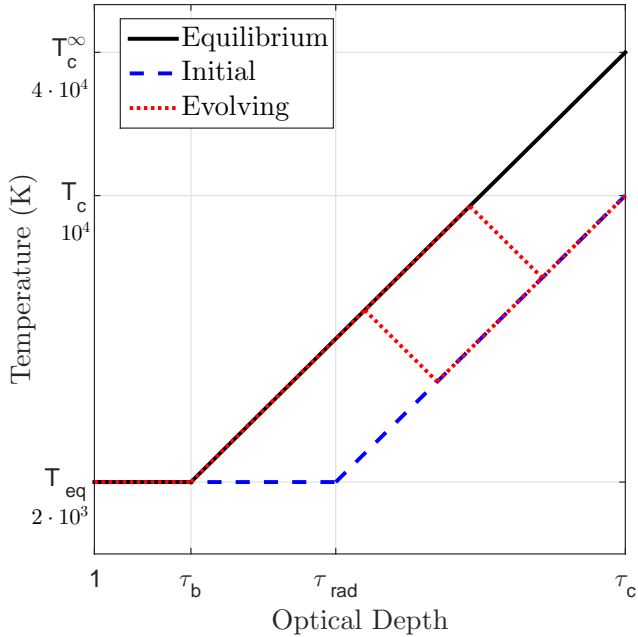}
\caption{Schematic temperature profile (logarithmic scale) of a hot Jupiter with an energy deposition that extends to its center. The equilibrium state (solid black line) is characterized by an equilibrium central temperature $T_c^\infty$. A hot-Jupiter profile with an initial $T_c<T_c^\infty$ (dashed blue line) is also plotted. Two intermediate stages (dotted red lines) show the reheating (re-inflation) of the planet from the initial to the equilibrium phase. Typical values of the temperature are provided.
\label{fig:reinflate}}
\end{figure}

It is clear from the discussion in Section \ref{sec:power_law} that the final equilibrium temperature profile of the planet is given by Figure \ref{fig:scheme_center}, with central temperature $T_c^\infty$, even if the planet had initially $T_c<T_c^\infty$ (the general equilibrium profile, in case the energy deposition does not reach the center, is given in Appendix \ref{sec:other_cases}). However, although planets cool down (and contract) or heat up (and expand) to the same temperatures (and radii), imposed by the stellar irradiation and the heat deposition, we show below that the timescales to reach equilibrium are different. In Figure \ref{fig:reinflate} we present a schematic plot of the reheating (and therefore re-inflation) of a planet from an initial central temperature $T_c$ (assuming the planet has cooled for a few Gyr in the absence of power deposition) to a final equilibrium central temperature $T_c^\infty>T_c$, imposed by the energy deposition. As seen in Figure \ref{fig:reinflate}, the planet heats up from the outside in. This result, which is also evident from numerical calculations by \citet{WuLithwick2013}, is due to both the increase in heat capacity and final temperature with depth and the decrease in deposition with depth, so the heating rate is $\propto\tau^{-(1+\alpha+\beta/4)}$. This outside-in heating implies that models, which exploit the entire heat deposited in the convective region to heat up the planet, overestimate the reheating rate \citep[see, e.g., the recent work by][who suggest a novel test to distinguish between reheating and stalling contraction]{LopezFortney2015}.

The time to heat the planet's center, which determines the re-inflation timescale is given by
\begin{equation}\label{eq:t_heat}
t_{\rm heat}\sim\frac{\frac{M}{m_p}k_{\rm B}T_c^\infty}{L_{\rm dep}(\tau_c)}.
\end{equation}
It is instructive to compare this timescale to the cooling timescale of initially hot planets to the same equilibrium central temperature. Due to the decrease of the internal luminosity with central temperature, the cooling timescale is also determined by the final equilibrium temperature $T_c^\infty$, as seen by combining Equations \eqref{eq:irradiated} and \eqref{eq:cooling}
\begin{equation}\label{eq:t_cool}
t_{\rm cool}\sim\frac{\frac{M}{m_p}k_{\rm B}T_c^\infty}{L_{\rm int}}=
\frac{\frac{M}{m_p}k_{\rm B}T_c^\infty}{L_{\rm dep}(\tau_{\rm rad})},
\end{equation}
where the last equality is due to the condition $L_{\rm int}=L_{\rm dep}(\tau_{\rm rad})$, which is fulfilled in the final cooling stage (see Section \ref{sec:power_law} and Figure \ref{fig:scheme_center}). By combining Equations \eqref{eq:t_heat} and \eqref{eq:t_cool} we find
\begin{equation}\label{eq:time_ratio}
\frac{t_{\rm heat}}{t_{\rm cool}}\sim\frac{L_{\rm dep}(\tau_{\rm rad})}{L_{\rm dep}(\tau_c)}=\left(\frac{\tau_c}{\tau_{\rm rad}}\right)^\alpha\approx 30,
\end{equation}  
with the numerical value calculated using $\tau_c=10^{10}$, $\tau_{\rm rad}=10^5$, and $\alpha=0.3$. Since the typical cooling time to a radius of $1.3R_J$ is $\sim 1\textrm{ Gyr}$ (see, e.g., Figure \ref{fig:power_law}), the long heating timescales of $\sim 30\textrm{ Gyr}$ (for $\epsilon=5\%$, which matches inflation to $1.3R_J$) imply that planets can only be mildly re-inflated with Ohmic dissipation (up to about $0.2 R_J$), and that they do not reach their final equilibrium temperature \citep[see][for similar results]{WuLithwick2013}. 

\section{Conclusions}
\label{sec:conclusions}

In this work we analytically studied the effects of additional power sources on the radius of irradiated giant gas planets. The additional heat sources halt the evolutionary cooling and contraction of gas giants, implying a large final equilibrium radius. A slowdown in the evolutionary cooling prior to equilibrium is possible only for sources which do not extend to the planet's center. 

We generalized our previous work \citepalias{GS2015}, which was confined to localized point sources, to treat sources that extend from the photosphere to the deep interior of the planet. We parametrized such a heat source by the total power it deposits below the photosphere $\epsilon L_{\rm eq}$, and by the logarithmic decay rate of the deposited power with optical depth $\alpha>0$. Implicitly, we assumed a heating profile $L_{\rm dep}(\tau)=\epsilon L_{\rm eq}\tau^{-\alpha}$, with $L_{\rm dep}(\tau)$ denoting the accumulated heat deposited below an optical depth $\tau$. Motivated by previous studies \citep[e.g.,][]{Batygin2011}, we measured the total heat with respect to the incident stellar irradiation $L_{\rm eq}$ and adopted the efficiency parameter $\epsilon$, which was assumed small $\epsilon <1$.

We generalized the technique used in \citetalias{GS2015} and showed that planetary cooling and contraction stop when the internal luminosity (i.e. cooling rate) drops below the heat deposited in the convective region $L_{\rm int}\lesssim L_{\rm dep}(\tau_{\rm rad})$ \citep[see also][]{WuLithwick2013}. This condition defines a threshold efficiency $\epsilon\gtrsim\tau_{\rm rad}^{-(1-\alpha)}$, required to explain the inflation of observed hot Jupiters, where $\tau_{\rm rad}\approx 10^5$ is the optical depth of the radiative-convective boundary (of $\sim 1$ Gyr old planets with an equilibrium temperature of $T_{\rm eq}\approx 2\cdot 10^3\textrm{ K}$) in the absence of heat deposition, and only flat enough heating profiles $\alpha<1$ have an impact. 

The method presented in this work reproduces previous numerical results while providing simple intuition and it may be used to study the effects of any power source on the radius and structure of hot Jupiters. Combining the model with observational correlations \citep[see, e.g.,][]{DemorySeager2011,Laughlin2011,MillerFortney2011,Schneider2011} may reveal the nature of the additional heat deposition mechanism, if exists, and provide a step toward solving the observed radius anomalies.

For a quantitative example, we focused on the suggested Ohmic dissipation mechanism, which stems from the interaction of atmospheric winds with the planet's magnetic field \citep[see, e.g.,][]{Batygin2011,HuangCumming2012,WuLithwick2013}.
This mechanism can be described by a power law $\alpha\approx 0.3$ in the planet's interior, and a reduction of $\approx 5\cdot 10^2$ in the efficiency, due to the slimness of the wind zone.
We therefore found that the threshold efficiency in this case is $\approx 5\%$, in accordance with previous numerical studies. 

Assuming a constant efficiency of 3\%, we estimated inflated radii which are similar to the numerical predictions of \citet{WuLithwick2013}. However, we challenged the assumption of a constant efficiency, made in previous studies, and examined the correlation between the efficiency and the equilibrium temperature. We found that the efficiency rises with temperature, due to the increase in electrical conductivity, to a maximum of $\approx 0.3$ at $T_{\rm eq}\approx 1500\textrm{ K}$, and then drops due to the magnetic drag \citep[see also][]{Menou2012}. As a result, we are able to explain the concentration of radius anomalies around this temperature \citep{Laughlin2011}, and to account for the radii of most inflated hot Jupiters, which are in the range $\approx 1500\textrm{ K}-2500\textrm{ K}$ and reach $\approx 1.6 R_J$. In addition, we argue that if these planets are indeed inflated by the Ohmic mechanism then they have already reached their final equilibrium state \citep[see also][]{Batygin2011}, and that the energy deposition must have suspended their contraction, and could not have re-inflated them from a smaller radius, since re-inflation timescales are too long \citep[see also][]{WuLithwick2013}. Nonetheless, some extremely inflated planets have radii which exceed the predictions of our model.   

In contrast to most previous studies, we did not introduce any free parameters to model the wind zone, but rather related the wind velocity, and therefore the amount of dissipated heat, to the strength of the magnetic field and to the equilibrium temperature. This procedure, combined with the generalized technique from \citetalias{GS2015}, enabled us to estimate the observational correlations expected in an Ohmic heating scenario, and to compare them with observations \citep{Laughlin2011,Schneider2011}. Although our main conclusions are robust, the exact shape of the radius-equilibrium-temperature curve should be studied with more detailed simulations, due to the approximate nature of our assessments. 

\acknowledgements
This research was partially supported by ISF, ISA and iCore grants. 
We thank Oded Aharonson, Konstantin Batygin, Peter Goldreich, Yohai Kaspi, Thaddeus Komacek, Yoram Lithwick, Adam Showman, and David J. Stevenson for insightful discussions. We also thank the anonymous referee for valuable comments, which improved the paper.  

\appendix

\section{General Power-Law Energy Deposition}
\label{sec:other_cases}

In Section \ref{sec:power_law} we analyzed the effects of extended heat deposition, parametrized by a power-law heating profile which extends from the surface to the interior of a hot Jupiter. However, in Section \ref{sec:power_law} we confined the discussion to heating profiles with a cumulative power index $\alpha<1-\beta$, which extend to the planet's center. In this section we relax both constrains, and address more general sources, which may have steeper profiles and a cut-off at some $\tau_{\rm cut}<\tau_c$. 

We now consider different values of $\alpha$, with a distinction made by the value of $\beta/(1-\alpha)$.

\begin{figure}[tbh]
\includegraphics[width=\columnwidth]{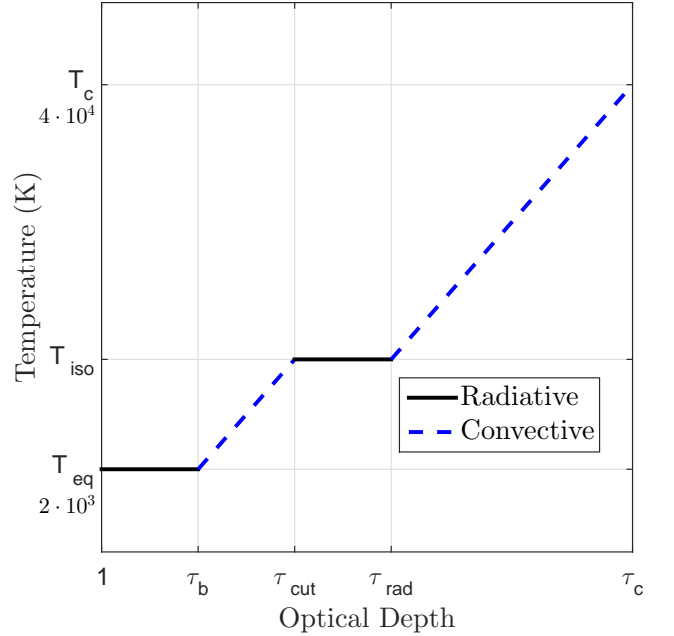}
\caption{Schematic temperature profile (logarithmic scale) of a hot Jupiter with an energy deposition that corresponds to {\it Case I}, i.e., $\alpha<1-\beta$ (see text). The structure of the planet is characterized by two radiative, nearly isothermal, regions (solid black lines) and two convective regions (dashed blue lines): the main convective interior, and an induced exterior secondary convective zone. Typical values of the temperature are provided.
\label{fig:scheme}}
\end{figure}

{\it Case I: $\alpha<1-\beta$}. In this case, as explained in Section \ref{sec:power_law}, a convective region emerges at $\tau_b$, given by Equation \eqref{eq:power_tau_b}. The convective region continues until $\tau_{\rm cut}$, reaching an energy density of
\begin{equation}\label{eq:power_u_eq}
\frac{U_{\rm iso}}{U_{\rm eq}}\sim\left(\frac{\tau_{\rm cut}}{\tau_b}\right)^\beta\sim\left(\tau_{\rm cut}\epsilon^{1/(1-\alpha)}\right)^\beta,
\end{equation}
which is derived similarly to Equations \eqref{eq:U_eq_eff_big} and \eqref{eq:u_final}. This secondary convective region is connected to the main interior convective region with a radiative tangent, and $L_{\rm int}$ is found using $U_{\rm iso}$ in the same manner as in Section \ref{sec:point}. A schematic temperature profile which clarifies the alternating radiative-convective structure, which is analogous to \citetalias{GS2015} and to Section \ref{subsec:deposition}, is given in Figure \ref{fig:scheme}.

{\it Case II: $1-\beta<\alpha<1$}. In this case there is no transition to a secondary convective region. Rather, the radiative profile of Equation \eqref{eq:u_tau_hat} continues up to $\tau_{\rm cut}$, reaching
\begin{equation}\label{eq:u_eq_eff_case2}
\frac{U_{\rm iso}}{U_{\rm eq}}\approx 1+\epsilon\tau_{\rm cut}^{1-\alpha}\approx\epsilon\tau_{\rm cut}^{1-\alpha},
\end{equation}
with the last approximation made for the significant heating regime.

{\it Case III: $\alpha>1$}. In this case, according to Equation \eqref{eq:rad_power}, the radiative profile is governed by low optical depths and
\begin{equation}
U=U_{\rm eq}+\frac{3}{\alpha-1}\frac{\epsilon L_{\rm eq}}{4\pi R^2c}
\end{equation}
for $\tau\gg 1$. It is easy to verify that there is no secondary convective region in this case for $\epsilon\ll 1$. Therefore, the radiation energy density of the deep isotherm is
\begin{equation}
\frac{U_{\rm iso}}{U_{\rm eq}}\approx 1+\epsilon\approx 1,
\end{equation}
regardless of $\tau_{\rm cut}$. We conclude that heating with $\epsilon\ll 1$ is unable to significantly effect the planetary cooling for $\alpha>1$.

\begin{figure}[tbh]
\includegraphics[width=\columnwidth]{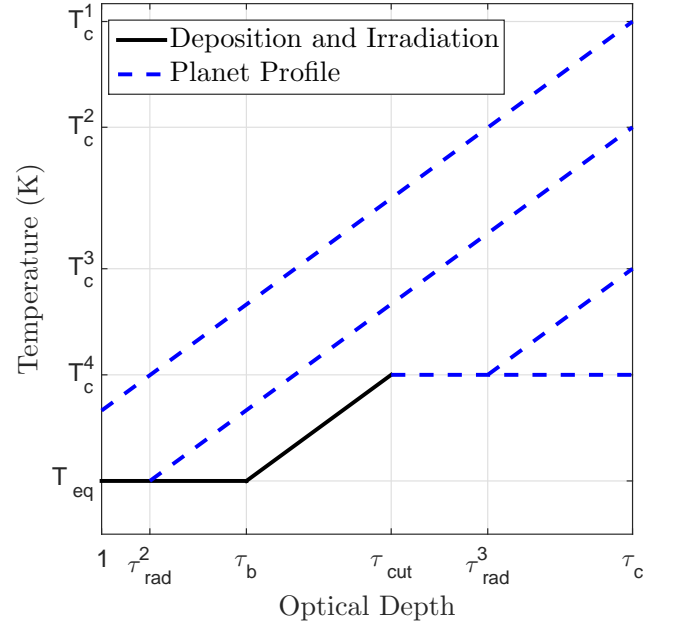}
\caption{Schematic temperature profiles (logarithmic scale) of hot Jupiters with an energy deposition with a cut-off at $\tau_{\rm cut}<\tau_c$. The lower bound on the outer temperature profile, set by a combination of the stellar irradiation and heat deposition (solid black line), is connected to the internal boundary condition (temperature $T_c$ at $\tau_c$). Planet profiles (dashed blue lines) are given for decreasing central temperatures, which correspond to the different evolutionary stages 1-4 in the text.
\label{fig:scheme_stages}}
\end{figure}

As in Section \ref{sec:point}, the decrease in the internal luminosity is given by
\begin{equation}\label{eq:l_u_eq}
\frac{L_{\rm int}}{L_{\rm int}^0}=\left(\frac{U_{\rm iso}}{U_{\rm eq}}\right)^{-(1-\beta)/\beta}.
\end{equation}
From both Equations \eqref{eq:power_u_eq} and \eqref{eq:u_eq_eff_case2} we find the critical criterion for a significant effect on the cooling rate of the planet, provided that $\alpha<1$:
\begin{equation}\label{eq:crit}
\epsilon\tau_{\rm cut}^{1-\alpha}\gtrsim 1.
\end{equation}
As discussed in Section \ref{sec:power_law}, and seen in Figure \ref{fig:scheme}, an additional requirement is that $\tau_b<\tau_{\rm rad}$ (for {\it Case I}, with a similar analog for {\it Case II}), with $\tau_{\rm rad}$ denoting the radiative-convective transition in the absence of heat deposition. This requirement is satisfied by the condition of Equation \eqref{eq:crit} for $\tau_{\rm cut}<\tau_{\rm rad}$. On the other hand, if the deposition is intense or deep enough, so that $U_{\rm iso}\sim U_c$, then the planet reaches equilibrium and cooling stops entirely, as explained in \citetalias{GS2015}. 

A more general analysis, presented schematically in Figure \ref{fig:scheme_stages}, shows that for a given heating profile with a cut-off at $\tau_{\rm cut}<\tau_c$, the planet evolves through 4 distinct stages, with Equation \eqref{eq:l_u_eq} relevant only to {\it Stage 3}:

{\it Stage 1: isolation.} For very high central temperatures, the planet is fully convective (since $\tau_{\rm rad}<1$), and its cooling rate is unaffected by the stellar irradiation or by the heat deposition \citepalias[see][]{GS2015}.

{\it Stage 2: irradiation.} As the planet cools, it develops a radiative envelope (which thickens with time) and its internal luminosity is determined by the stellar irradiation (see Section \ref{subsec:irradiation}) but is unaffected by the heat deposition (since $\tau_{\rm rad}<\tau_b$).

{\it Stage 3: deposition.} At even lower central temperatures (when $\tau_{\rm rad}>\tau_b$), the planet's cooling rate is reduced by the heat deposition, according to Equation \eqref{eq:l_u_eq}. This stage is also depicted in Figure \ref{fig:scheme}.

{\it Stage 4: equilibrium.} When the planet reaches $T_c=T_{\rm iso}$, evolutionary cooling stops entirely, and the planet reaches its final state.

In the special case of a heating profile without a cut-off ($\tau_{\rm cut}=\tau_c$), the planet skips the intermediate {\it Stage 3} and transitions directly from {\it Stage 2} (cooling unaffected by deposition) to the equilibrium state. This transition is explained in Section \ref{sec:power_law}, and is evident from Figures \ref{fig:scheme_center} and \ref{fig:scheme_stages}.
Essentially, the equilibrium central temperature $T_c^\infty$, imposed by the heat deposition and introduced in Section \ref{sec:power_law}, is a special case (for $\tau_{\rm cut}=\tau_c$) of the deep isotherm temperature $T_{\rm iso}$.

Combining Equations \eqref{eq:irradiated}, \eqref{eq:cooling}, and \eqref{eq:l_u_eq}, we find the effect of heating on the central temperature (and radius) at a given age, during the relevant {\it Stage 3}, when the cooling is influenced by the heat deposition
\begin{equation}\label{eq:t_c}
\Delta R(t)\propto T_c(t)\propto\left(\frac{U_{\rm iso}}{U_{\rm eq}}\right)^{(1-\beta)/(4-\beta-\beta b)}\approx\left(\frac{U_{\rm iso}}{U_{\rm eq}}\right)^{0.5},
\end{equation}
with $\beta=0.35$ and $b=7$ estimated in \citetalias{GS2015}, and with the ratio $U_{\rm iso}/U_{\rm eq}$ given by Equation \eqref{eq:power_u_eq} or \eqref{eq:u_eq_eff_case2}.

\section{High Rossby Number Regime}
\label{sec:equator}

In Section \ref{subsec:winds_currents} we adopted a low Rossby number approximation ${\rm Ro}\ll 1$, in which the advective term $v\nabla v\sim v^2/R$ in the force balance equation is negligible when compared to the Coriolis acceleration $2\Omega v\sin\phi$. Although it is justified for the atmosphere on average, this approximation breaks down close to the equator ($\phi=0$). In this section we reanalyze the atmospheric wind velocity derivation in the high Rossby number limit ${\rm Ro}\gg 1$, relevant for the equator, and compare the results to the conclusions of Section \ref{subsec:winds_currents}.
	
In the high $\rm{Ro}$ case, the force balance Equation \eqref{eq:wind_vel} is replaced with
\begin{equation}\label{eq:wind_vel_high_ro}
\frac{c_s^2}{R}\frac{\Delta T}{T_{\rm eq}}=\Omega v\left(\frac{v}{\Omega R}+\frac{\sigma}{\sigma_c}\frac{\rho_c}{\rho}\right),
\end{equation}
where the advective term replaces the (now negligible) Coriolis term. Equation \eqref{eq:wind_vel_high_ro} indicates that $v<c_s$, and therefore ${\rm Ro}=v/(2\Omega R\sin\phi)<1$, except for the equator (since $c_s\sim\Omega R$; see Section \ref{subsec:winds_currents}). This understanding, together with a similar insight from the low Rossby number Equation \eqref{eq:wind_vel}, self-consistently justifies our low Ro approximation for the atmosphere on average (excluding the equator). We see again, as in the low Ro case, that due to the strong dependence of the conductivity on the temperature, the magnetized regime corresponds to high equilibrium temperatures. In addition, Equation \eqref{eq:epsilon_power_law_sigma}, which demonstrates the dependence of the heating efficiency $\epsilon\propto\sigma v^2$ on the conductivity (the dominant parameter for an intuitive understanding, due to its strong dependence on the temperature), is clearly valid in the high Ro limit as well, so $\epsilon\propto\sigma$ for low conductivities, while $\epsilon\propto\sigma^{-1}$ for high conductivities \citep[see also][]{Menou2012}.

The decay of the velocity and day-night temperature difference with depth is calculated similarly to Section \ref{subsec:winds_currents}, with Equation \eqref{eq:wind_vel_high_ro} replacing Equation \eqref{eq:wind_vel}. By combining Equation \eqref{eq:wind_vel_high_ro} with the diffusion Equation \eqref{eq:diffusion}, we find that for the unmagnetized regime (the magnetized regime is indifferent to Ro)
\begin{equation}\label{eq:delta_t_high_ro}
\frac{v}{c_s}=\left(\frac{\Delta T}{T_{\rm eq}}\right)^{1/2}=\frac{R\sigma_{\rm SB}T_{\rm eq}^4}{c_s^3}\frac{\kappa}{\tau^2}.
\end{equation}
Equation \eqref{eq:delta_t_high_ro}, which is the high Ro version of Equation \eqref{eq:delta_t}, shows that the velocity and temperature difference decay with depth in the high Ro regime as well. Quantitatively, we find that the decay of the velocity with depth $v\propto\kappa/\tau^2$ is the same in both regimes (low and high Ro), as can be immediately understood from Equation \eqref{eq:diffusion}.

We conclude that the high Ro regime, relevant for a narrow strip around the equator, exhibits the same qualitative behavior as the low Ro regime, with some of the quantitative results reproduced as well. Nevertheless, some of the specific power-law scalings with the equilibrium temperature change in this regime. 

\bibliographystyle{apj}

\begin{thebibliography}

\bibitem[Allard et al.(2001)]{Allard2001} Allard, F., Hauschildt, P.~H., Alexander, D.~R., Tamanai, A., \& Schweitzer, A.\ 2001, \apj, 556, 357

\bibitem[Anderson et al.(2011)]{Anderson2011} Anderson, D.~R., Smith, A.~M.~S., Lanotte, A.~A., et al.\ 2011, \mnras, 416, 2108

\bibitem[Arras \& Bildsten(2006)]{ArrasBildsten2006} Arras, P., \& Bildsten, L.\ 2006, \apj, 650, 394

\bibitem[Arras \& Socrates(2009a)]{ArrasSocrates2009a} Arras, P., \& Socrates, A.\ 2009a, arXiv:0901.0735

\bibitem[Arras \& Socrates(2009b)]{ArrasSocrates2009b} Arras, P., \& Socrates, A.\ 2009b, arXiv:0912.2318

\bibitem[Arras \& Socrates(2010)]{ArrasSocrates2010} Arras, P., \& Socrates, A.\ 2010, \apj, 714, 1

\bibitem[Baraffe et al.(2010)]{Baraffe2010} Baraffe, I., Chabrier, G., \& Barman, T.\ 2010, RPPh, 73, 016901

\bibitem[Baraffe et al.(2003)]{Baraffe2003} Baraffe, I., Chabrier, G., Barman, T.~S., Allard, F. \& Hauschildt, P.~H.\ 2003, \aap, 402, 701

\bibitem[Baraffe et al.(2014)]{Baraffe2014} Baraffe, I., Chabrier, G., Fortney, J., \& Sotin, C.\ 2014, Protostars and Planets VI (Tucson: Univ. Arizona Press), 763

\bibitem[Batygin \& Stevenson(2010)]{BatyginStevenson2010} Batygin, K., \& Stevenson, D.~J.\ 2010, \apjl, 714, 238

\bibitem[Batygin et al.(2011)]{Batygin2011} Batygin, K., Stevenson, D.~J., \& Bodenheimer, P.~H.\ 2011, \apj, 738, 1

\bibitem[Bodenheimer et al.(2003)]{Bodenheimer2003} Bodenheimer, P., Laughlin, G., \& Lin, D.~N.~C.\ 2003, \apj, 592, 555

\bibitem[Bodenheimer et al.(2001)]{Bodenheimer2001} Bodenheimer, P., Lin, D.~N.~C., \& Mardling, R.~A.\ 2001, \apj, 548, 466

\bibitem[Burrows et al.(2000)]{Burrows2000} Burrows, A., Guillot, T., Hubbard, W.~B., et al.\ 2000, \apjl, 534, L97

\bibitem[Burrows et al.(2007)]{Burrows2007} Burrows, A., Hubeny, I., Budaj, J., \& Hubbard, W.~B.\ 2007, \apj, 661, 502

\bibitem[Burrows et al.(1997)]{Burrows97} Burrows, A., Marley, M., Hubbard, W.~B., et al.\ 1997, \apj, 491, 756

\bibitem[Chabrier \& Baraffe(2007)]{ChabrierBaraffe2007} Chabrier, G., \& Baraffe, I.\ 2007, \apjl, 661, L81

\bibitem[Chabrier et al.(2004)]{Chabrier2004} Chabrier, G., Barman, T., Baraffe, I., Allard, F., \& Hauschildt, P.~H.\ 2004, \apjl, 603, L53

\bibitem[Chan et al.(2011)]{Chan2011} Chan, T., Ingemyr, M., Winn, J.~N., et al.\ 2011, \apj, 141, 179

\bibitem[Christensen et al.(2009)]{Christensen2009} Christensen, U.~R., Holzwarth, V., \& Reiners, A.\ 2009, \nat, 457, 167

\bibitem[Demory \& Seager(2011)]{DemorySeager2011} Demory, B., \& Seager, S.\ 2011, \apjs, 197, 12

\bibitem[Fortney \& Nettelmann(2010)]{FortneyNettelmann2010} Fortney, J.~J., \& Nettelmann, N.\ 2010, \ssr, 152, 423

\bibitem[Freedman et al.(2008)]{Freedman2008}
Freedman, R.~S., Marley, M.~S., \& Lodders, K.\ 2008, \apjs, 174, 504. 

\bibitem[Ginzburg \& Sari(2015)]{GS2015} Ginzburg, S. \& Sari, R.\ 2015, \apj, 803, 111

\bibitem[Gu et al.(2003)]{Gu2003} Gu, P., Lin, D.~N.~C., \& Bodenheimer, P.~H.\ 2003, \apj, 588, 509

\bibitem[Guillot et al.(1996)]{Guillot96} Guillot, T., Burrows, A., Hubbard, W.~B., Lunine, J.~I., \& Saumon, D.\ 1996, \apjl, 459, L35

\bibitem[Guillot \& Showman(2002)]{GuillotShowman2002} Guillot, T., \& Showman, A.~P.\ 2002, \aap, 385, 156

\bibitem[Hartman et al.(2011)]{Hartman2011} Hartman, J.~D., Bakos, G.~\`{A}., Torres, G., et al.\ 2011, \apj, 742, 59

\bibitem[Huang \& Cumming(2012)]{HuangCumming2012} Huang, X., \& Cumming, A.\ 2012, \apj, 757, 47

\bibitem[Ibgui \& Burrows(2009)]{IbguiBurrows2009} Ibgui, L., \& Burrows, A.\ 2009, \apj, 700, 1921

\bibitem[Ibgui et al.(2010)]{Ibgui2010} Ibgui, L., Burrows, A., \& Spiegel, D.~M.\ 2010, \apj, 713, 751

\bibitem[Ibgui et al.(2011)]{Ibgui2011} Ibgui, L., Spiegel, D.~M., \& Burrows, A.\ 2011, \apj, 727, 75

\bibitem[Iro et al.(2005)]{Iro2005} Iro, N., B{\'e}zard, B., \& Guillot, T.\ 2005, \aap, 436, 719

\bibitem[Jackson et al.(2008)]{Jackson2008} Jackson, B., Greenberg, R., \& Barnes, R.\ 2008, \apj, 681, 1631

\bibitem[Kammer et al.(2015)]{Kammer2015} Kammer, J.~A., Knutson, H.~A, Line, M.~R., et al. 2015, \apj, 810, 118

\bibitem[Knutson et al.(2009)]{Knutson2009} Knutson, H.~A, Charbonneau, D,  Cowan, N.~B., et al. 2009, \apj, 690, 822

\bibitem[Komacek \& Showman(2015)]{KomacekShowman2015} Komacek T.~D., \& Showman, A.~P.\ 2015, arXiv:1601.00069

\bibitem[Laughlin et al.(2011)]{Laughlin2011} Laughlin, G., Crismani, M., \& Adams, F.~C.\ 2011, \apjl, 729, L7

\bibitem[Leconte \& Chabrier(2012)]{LeconteChabrier2012} Leconte, J., \& Chabrier, G.\ 2012, \aap, 540, A20

\bibitem[Leconte et al.(2010)]{Leconte2010} Leconte, J., Chabrier, G., Baraffe, I., \& Levard, B.\ 2010, \aap, 516, A64

\bibitem[Liu et al.(2008)]{Liu2008} Liu, X., Burrows, A., \& Ibgui, L.\ 2008, \apj, 687, 1191

\bibitem[Lopez \& Fortney(2015)]{LopezFortney2015} Lopez, E.~D., \& Fortney, J.~J.\ 2015, arXiv:1510.00067

\bibitem[Marleau \& Cumming(2014)]{MarleauCumming2014} Marleau, G.~D., \& Cumming, A.\ 2014, \mnras, 437, 1378

\bibitem[Menou(2012)]{Menou2012} Menou, K.\ 2012, \apj, 745, 138

\bibitem[Miller \& Fortney(2011)]{MillerFortney2011} Miller, N., \& Fortney, J.~J.\ 2011, \apjl, 736, L29

\bibitem[Miller et al.(2009)]{Miller2009} Miller, N., Fortney, J.~J., \& Jackson, B.\ 2009, \apj, 702, 1413

\bibitem[Peebles(1964)]{Peebles64} Peebles, P.~J.~E.\ 1964, \apj, 140, 328

\bibitem[Perez-Becker \& Showman(2013)]{PBS2013} Perez-Becker, D., \& Showman, A.~P.\ 2013, \apj, 776, 134

\bibitem[Perna et al.(2012)]{Perna2012} Perna, R., Heng, K., \& Pont, F.\ 2012, \apj, 751, 59

\bibitem[Perna et al.(2010)]{Perna2010} Perna, R., Menou, K., \& Rauscher, E.\ 2010, \apj, 719, 1421

\bibitem[Rauscher \& Menou(2013)]{RauscherMenou2013} Rauscher, E., \& Menou K.\ 2013, \apj, 764, 103

\bibitem[Rogers \& Komacek(2014)]{RogersKomacek2014} Rogers, T.~M., \& Komacek T.~D.\ 2014, \apj, 794, 132

\bibitem[Rogers \& Showman(2014)]{RogersShowman2014} Rogers, T.~M., \& Showman A.~P.\ 2014, \apjl, 782, L4

\bibitem[Saumon et al.(1995)]{Saumon95} Saumon, D., Chabrier, G., \& Van Horn, H.~M.\ 1995 \apjs, 99, 713

\bibitem[Schneider et al.(2011)]{Schneider2011} Schneider, J., Dedieu, C., Le Sidaner, P., Savalle, R., \& Zolotukhin, I.\ 2011, \aap, 532, A79

\bibitem[Showman et al.(2010)]{Showman2010} Showman, A.~P., Cho, J.~Y.-K., \& Menou, K.\ 2010, in Exoplanets, ed. S. Seager
(Space Science Series; Tucson, AZ: Univ. Arizona Press), 471

\bibitem[Showman et al.(2008)]{Showman2008} Showman, A.~P., Cooper, C.~S., Fortney, J.~J, \& Marley, M.~S.\ 2008, \apj, 682, 559

\bibitem[Showman \& Guillot (2002)]{ShowmanGuillot2002} Showman, A.~P., \& Guillot, T.\ 2002, \aap, 385, 166

\bibitem[Socrates(2013)]{Socrates2013} Socrates, A.\ 2013, arXiv:1304.4121

\bibitem[Spiegel \& Burrows(2012)]{SpiegelBurrows2012} Spiegel, D.~S., \& Burrows, A.\ 2012, \apj, 745, 174

\bibitem[Spiegel \& Burrows(2013)]{SpiegelBurrows2013} Spiegel, D.~S., \& Burrows, A.\ 2013, \apj, 772, 76

\bibitem[Winn \& Holman(2005)]{WinnHolman2005} Winn, J.~N., \& Holman, M.~J.\ 2005, \apjl, 159, L159

\bibitem[Wong et al.(2015)]{Wong2015} Wong, I., Knutson, H.~A., Lewis, N.~K., et al.\ 2015, \apj, 811, 122

\bibitem[Wu \& Lithwick(2013)]{WuLithwick2013} Wu, Y., \& Lithwick, Y.\ 2013, \apj, 763, 13

\bibitem[Youdin \& Mitchell(2010)]{YoudinMitchell2010} Youdin, A.~N., \& Mitchell, J.~L.\ 2010, \apj, 721, 1113

\end{thebibliography}

\end{document}